\documentstyle[12pt]{article}

\newcommand{\bsp}{\begin{sloppypar}}
\newcommand{\esp}{\end{sloppypar}}
\newcommand{\ds}{\displaystyle}
\newcommand{\la}[1]{\label{#1}}
\newcommand{\re}[1]{\ (\ref{#1})}
\newcommand{\nn}{\nonumber}
\newcommand{\ed}{\end{document}}
\newcommand{\be}{\begin{equation}}
\newcommand{\ee}{\end{equation}}
\newcommand{\ba}{\begin{eqnarray}}
\newcommand{\ea}{\end{eqnarray}}
\newcommand{\bb}{\begin{thebibliography}}
\newcommand{\eb}{\end{thebibliography}}
\newcommand{\bi}[1]{\bibitem{#1}}
\newcommand{\ct}[1]{\cite{#1}}

\newcommand{\th}{theoretical}
\newcommand{\ps}{pseudoscalar}
\newcommand{\sg}{strange}
\newcommand{\il}{integral}
\newcommand{\nr}{nonrelativistic}
\newcommand{\pth}{perturbation theory}
\newcommand{\pto}{proportional to}
\newcommand{\La}{Lagrangian}
\newcommand{\pr}{parameter}
\newcommand{\op}{operator}
\newcommand{\fq}{frequency}
\newcommand{\fl}{flavour}
\newcommand{\tsi}{takes into account}
\newcommand{\tia}{take into account}
\newcommand{\ten}{taken into account}
\newcommand{\ft}{fluctuation}
\newcommand{\oge}{one-gluon exchange}
\newcommand{\dep}{dependen}
\newcommand{\mm}{magnetic moment}
\newcommand{\cl}{calculate}
\newcommand{\fw}{in the framework of}
\newcommand{\pg}{propagator}
\newcommand{\rl}{relativistic}
\newcommand{\ki}{kinetic}
\newcommand{\po}{potential}
\newcommand{\mlp}{multiplet}
\newcommand{\qt}{quantitative}
\newcommand{\ph}{phenomenolog}
\newcommand{\elm}{electromagnetic}
\newcommand{\sy}{symmetr}
\newcommand{\spl}{splitting}
\newcommand{\gs}{ground state}
\newcommand{\bc}{boundary condition}
\newcommand{\bm}{bag model}
\newcommand{\cf}{confinement}
\newcommand{\cst}{constituent}
\newcommand{\con}{contribution}
\newcommand{\cu}{current}
\newcommand{\di}{difference}
\newcommand{\dc}{distance}

\newcommand{\ef}{effect}
\newcommand{\ex}{experiment}
\newcommand{\fu}{function}
\newcommand{\hn}{hadron}

\newcommand{\Hw}{However}
\newcommand{\s}{instanton}
\newcommand{\ia}{interaction}

\newcommand{\np}{nonperturbative}

\newcommand{\p}{proton}
\newcommand{\q}{quark}

\newcommand{\vm}{vacuum}
\newcommand{\wf}{wave function}
\newcommand{\w}{where}

\newcommand{\cd}{condensate}
\newcommand{\qm}{quark model}
\newcommand{\pl}{particle}
\newcommand{\str}{structure}
\begin{document}
{\hfill     preprint JINR E2-94-231}

{\hfill     June, 1994}\\
\begin{center}
{\bf
ELECTROMAGNETIC HADRON MASS DIFFERENCES AND ESTIMATION OF ISOTOPIC SYMMETRY
VIOLATION PARAMETERS OF QCD VACUUM FROM QUARK MODEL.
}\\[5mm]
A. E. DOROKHOV
\footnote{\large {
E-mail:dorokhov@thsun1.jinr.dubna.su}}\\
{\small \it Joint Institute for Nuclear Research,} \\
{\small \it Bogoliubov Theoretical Laboratory,} \\
{\small \it Head Post Office P. O. Box 79,SU-101000 Moscow, Russia}\\[13mm]

Submitted to {\it Nucl.Phys. A}\\[13mm]

{\bf Abstract} \\
\end{center}

The estimations of the light quark mass differences, $m_d-m_u$, and
the light quark condensate differences, $<\bar DD> - <\bar UU>$,
are obtained in the framework of the \qm\ with QCD vacuum induced \q\ \ia .
We consider long - wave condensate and short - wave \s\ contributions
to the electromagnetic \hn\ mass \di s and
show that the latter significantly improve the results on baryon octet.
The results are: $m_d-m_u \approx 3.5\ MeV$ and
$<\bar DD> - <\bar UU> \approx -(0\sim 3)\cdot10^{-3}<\bar UU>$.
\vspace{1.0cm}
\newpage

To describe the mass spectrum of \hn ic \gs s, it is actually necessary to
define two quantities: the scale of \hn ic masses and the scale of spin-spin
\spl . Then the \hn\ spectrum is almost reproduced by using the $SU_f(3)$
\sy y. This is the reason for success of most models of \hn s.
However, there exist more subtle \ef s such as, for example,
\elm\ mass differences (EMD) or the mass
spectrum of excited  and exotic states which as a
whole are a good base to select the most adequate approach to the \ph y of
strong \ia\ at low energies.

At present, the possibility of determination of the isospin symmetry violation
of light \q\ masses and their \cd s is widely discussed within different
approaches
\ct{HatsudaEMD}-\nocite{BrownEMD,IoffeEMD,EletIof,NasonEMD,Kissl}\ct{JinNiel}.
Mainly, the interest in this
problem is based on the necessity to relate the isotopic symmetry violation
on the level of \hn s to the differences of some intimate QCD parameters:
masses and condensates of light quarks. This intriguing problem is well known
for many years and is due to
the absence of \qt\ understanding of the QCD at low energies.
During the past decade the data for \hn\ mass \di s
have become much more accurate \ct{DPG}.
For the joined treatment of the \hn\ masses and their isospin \spl s
based on QCD inspired approaches
(QCD sum rules, Quark models) the important question is to determine, from
the \hn\ spectrum, the magnitudes of isospin \sy y violation in the light
\q\ masses and \cd s.

Another point is that the isospin symmetry breaking \ef s are tightly
related with charge \sy y breaking phenomena in nuclear physics.
Understanding of the latter is very important for a more profound view on
strong interaction forces \ct{CSBrev}.

On the one hand, EMD are just of \elm\ nature (in the presence of strong \ia )
and, in principle, calculable in terms of what is known. On
the other hand, the \ef s that strong \ia\ creates can perhaps be avoided by
using adequate approaches in calculations (e.g. Quark models, QCD sum rules,
Lattice QCD).
The mass \di s between members of the same isospin \mlp\
are due to two reasons: the proper \elm\ \ia\ between different \q s in
a \hn , $\Delta E_{\ds em}$, and the self-\ia\ of the \q s themselves.
The last one produces a \di\ between $u-$ and $d-$ \q\ masses,
$\Delta m=m_d-m_u$, which results in the
dependence of the strong \ia\ \po , $\Delta E_{\ds strong}$,
and \q\ \ki\ energy,  $\Delta E_{\ds kin}$, on $\Delta m$.

A consistent consideration of the \elm\ \ia\ of \q s in the presence
of strong \ia\ is possible within the \rl\ \bm\ \ct{Bo,mit2}.
It allows, in principle,
to calculate the self-energy \ia\ and determine the
$u-$ and $d-$ \q\ masses as the poles of the \q\ \pg\ in the bag.
\Hw , this problem is not fully solved yet \ct{Viollier}. So, we shall
calculate the \ia s between \q s with the value of the \q\ mass
\di\ fixed by fitting to the \ex al values of EMD.

In \ct{Desh}, \fw\ the MIT \bm\ \ct{mit2},
the \elm\ \ia\ between different \q s in \gs\ \hn s, $\Delta E_{\ds em}$,
has been \cl d. It has been shown that EMD are much more sensitive to
\q -\q\ \wf\ correlations than the masses and \mm s of \hn\ \gs s and are
a strong testing of the model. Further in \ct{BTh},  the \dep ce of
the \oge\ \po , $\Delta E_{\ds gl}$, on \q\ masses has been taken into account.
The $u - d$ \q\ mass \di\ has been estimated, $\Delta m \approx
4\ MeV$. It agrees with the current algebra estimation \ct{Leutw}.
In \ct{DKEM1}, the \ef s of \s s and \q\ \cd s on iso\mlp\ mass \spl\ of
baryons has been considered. It has been shown that these \con s
systematically improve the results for $\Sigma$ and $\Xi$ baryons. The
important role of \s s for baryon octet splittings has been noted in
\ct{YWL}, too.

In the present paper, we shall consider the isospin mass splittings of low -
lying
\hn s and obtain estimations of  the isospin violation in \q\ masses and \cd s.
To this end, we shall use the version of the bag model
based on the idea that the \ia\ of \hn\ \cst s
with background \vm\ fields in the bag plays the dominant role \ct{DKbm} .
It has been shown that the spin - dependent forces are determined by the \ia\
of \q s with \s s (short - range \vm\ \ft )
\ct{DKbm}-\nocite{Koch}\ct{ShurRos},
while the stability (\cf\ forces) is due to their \ia\ with \cd s (long -
wave \vm\ \ft ) \ct{DKbm,Hanss}.
This \np\ \ia\ between \q s strongly depending on \q\ masses
defines the spectroscopy of the ground states of \hn s. The results obtained
agree well with the \ex al ones. In addition to the assumption made in
\ct{DKEM1}, we shall \tia\ the center - of - masses and
gluon \cd\
corrections to \hn\ masses and allow the $SU(2)$ violation of light \q\ \cd s.

The main assumption of the model that the \ia\ of \q s and
gluons localized in the bag (thus at \ef ively small \dc s) with background
QCD \vm\ fields defines the \hn\ \str\ is analogous to the QCD sum rule
ideas \ct{SVZ,RRYa}. In the latter case the correlator of \hn\ currents is a
nonlocal object selecting the lowest \hn\ states.
Here, one also suggests implicitly that
local sources do not perturb the properties of physical \vm , i.e. the
values of \q\ and gluon \cd s.
In our case the extended bag plays a role similar to a \cu\ correlator
within the QCD sum rule.

Let us consider  the total
energy \di\ between two members of a \mlp . It is given by the expression:
\be
\Delta M_{tot}= \left\{\Delta E_{tot}-
{\ds \Delta <P^2>\over\ds 2E_{tot}}\right\}
{\ds E_{tot}\over\ds M_{tot}}
\la{DMtot}\ee
with
\be
\Delta E_{\ds tot} =
\Delta E_{\ds kin} + \Delta E_{\ds em}+ \Delta E_{\ds strong},
\la{Eem}\ee
and
\ba
\Delta E_{\ds strong} =
\Delta E_{\ds vac}+ \Delta E_{\ds inst} + \Delta E_{\ds gl},
\nn\ea
\w\ $M_{tot}$ is a \hn\ mass, $M_{tot}^2=E_{tot}^2-<P^2>$, with
center - of - mass motion correction $<P^2>$ \ten\ \ct{DJ},
$\Delta E_{\ds gl}$ is due to the QCD hyperfine \ia\ of \q s inside a bag,
and $\Delta E_{\ds vac}$ and $\Delta E_{\ds inst}$ are due to the
\ia\ of \q s with \vm\ fields.

 Detailed calculations of the \con s $\Delta E_{\ds em}$, $\Delta E_{\ds kin}$
and
 $\Delta E_{\ds gl}$ have been carried out in \ct{Desh,BTh}.
These \con s for two members $A$ and $B$ of a \mlp\ are given by the
expressions:
\ba
&&\Delta E_{kin}={\ds 1\over\ds R}
\sum_{i}^{u,d,s}(N_i^A-N_i^B)\omega(m_iR), \la{dEkin}\\
&&\Delta E_{gl}={\ds \alpha_s\over\ds 4R}
\sum_{i>j}^{u,d,s}(N_{ij}^A-N_{ij}^B)M_{gl}^{ij}I_{gl}(m_iR,m_jR), \la{dEgl}\\
&&\Delta E_{em}={\ds \alpha \over\ds R}
\sum_{i>j}^{u,d,s}(N_{ij}^A-N_{ij}^B)M_{em}^{ij}I_{em}(m_iR,m_jR),
\la{dEel}\ea
where $N_i$ is the number of light \q s of the \fl\ $i$ in the \hn ,
$N_{ij}$ is the number of light \q\ pairs in a \hn ,
$m_i$ is the \cu\ \q\ mass, $\omega_i$ is the mode \fq , $M^{ij}$ are averaged
over the \hn\ state (color -) spin \op , $I$ are
strengths and $\alpha$ are couplings for gluon and photon \ia s,
respectively. (Our definition of $\alpha_s$ differs from that used in \ct{mit2}
by factor $4$ and corresponds to the standard definition used in QCD.)

It is important that in the framework of the \bm\
$\Delta E_{em}$ is calculated explicitly, with no free \pr s. Contrary to the
\bm , these values have been not determined within the QCD sum rules
\ct{IoffeEMD,EletIof} and so this method is not completely self - consistent
in the determination of isotopic hadron mass differences.

The \con s, $\Delta E_{vac}$ and $\Delta E_{inst}$, are discussed in detail
in \ct{DKbm}.
The first term is caused by the interaction of \q s with low - \fq\ \vm\ fields
which gives the \cf\ of \q s. The \ia\ \La\ is expressed by:
\be
\Delta{\cal L}^{vac}=[\bar q(x)\Theta_V(x)]({i\over2}
\stackrel{\leftrightarrow}{\hat\partial}-m)Q(x)+
\bar Q(x)({i\over2}
\stackrel{\leftrightarrow}{\hat\partial}-m)[q(x)\Theta_V(x)]+
g\bar q(x)\gamma^\mu{\lambda^a\over2} q(x){\cal A}^a_\mu(x)\Theta_V(x),
\la{Lvac}\ee
Here the anticommuting external \q\ field $Q$ and
external gauge field ${\cal A}^a_\mu(x)$  are the \vm\ solution of QCD
equations parametrized by the values of their \cd s, and the localized \q\
field
$q(x)$ is given by the solution of the Dirac equation in a static spherical
cavity of radius $R$; the \fu\ $\Theta_{\ds v}(x)$  defines the
volume $V$ of the bag.  Expression \re{Lvac}
follows directly from the QCD (bag) \La\ by singling a \vm\ component out
of the \q\ field: $\Psi(x)=q(x)\Theta_V(x)+Q(x)$ in analogy with the procedure
used in the QCD sum rule technique \ct{SVZ}. It is supposed that these
components weakly correlate with each other (see the discussion on hierarchy of
\vm\ and \cst\ fields in \ct{DKbm}c).

Vacuum \cd\ induced corrections to the \hn\ mass are calculated by using the
stationary \pth\  with the \ia\ \La\ \re{Lvac}.
The resulting formulas are:
\ba
&&E^{QQ}_{vac}=-\left(\sum_{i}^{u,d,s}N_i A^{QQ}_i<0| \bar Q_i Q_i |0>\right)
 R^2+...,
\la{EvacQQ}\\
&&E^{GG}_{vac}=
\left(\sum_{i}^{u,d,s}N_i A_i^{GG}\right)
<0|{\ds \alpha_s\over\ds \pi}G^a_{\mu\nu}G^{a\mu\nu}|0> R^3+...,
\la{EvacGG}\ea
with
\[
A^{QQ}_i = \frac{\pi}{12} {\ds (\omega_i+m_iR)^2\omega_i
\over\ds \xi_i^2[2\omega_i(\omega_i - 1)+m_iR]}         ,\ \ \ \ \
A^{GG}_i =   {\ds \pi^2\over\ds 144}I^{GG}(m_iR),
\]
\w\ $<0| \bar Q_i Q_i |0>$ is the \q\ \cd\ of the $i$-th \fl ,
$<0|{\ds \alpha\over\ds \pi}G^a_{\mu\nu}G^{a\mu\nu}|0>$ is the gluon \cd ,
$\omega_i=(\xi_i^2+m_i^2R^2)^{1/2}$ is the one - \pl\ \fq ,
and $\xi_i$ is determined from the solution of the equation
originating from the bag \bc :
\be
\tan \xi_i=\xi_i/[1-m_iR-(\xi_i^2+m_i^2R^2)^{1/2}],
\la{egval}\ee
$I^{GG}(m_iR)$ is a \fu\ of masses with, for instance, $I^{GG}(0)=0.124$
and $I^{GG}(m_s R= 1.1)=0.130$. The calculations of $E^{QQ}$ and $E^{GG}$
are carried out in a fixed - point gauge \ct{RRYa}.
Dots in \re{EvacQQ} and \re{EvacGG} mean the \con s of \cd s of higher
dimensions which are numerically suppressed \ct{DKbm}.

 From \re{EvacQQ} in the first order of expansion in the small \q\ mass
\pr\ $m_q$ and the \di\
$\gamma=\frac{\ds <\bar DD> - <\bar UU>}{\ds <\bar UU>},\ (\gamma<0)$,
we obtain an increase of the \hn\ mass by:
\ba
\Delta E_{\ds vac}=&&-<\bar UU>R^2 \{B^{QQ}
\sum_{\ds i=u,d}N_{\ds i}m_{\ds i}
+\frac{\ds \pi}{24(\xi_0-1)}N_{\ds d}\gamma\} \nn\\
&&+< {\ds \alpha_s\over\ds \pi}G^2> R^3 B^{GG}
\sum_{\ds i=u,d}N_{\ds i}m_{\ds i},
\la{dEvac}\ea
\w\
$B^{QQ} = \left({\ds \partial A^{QQ}\over\ds \partial m_q}\right)_{\ds m_q=0}
\approx 0.202\ GeV^{-1}$ and
$B^{GG} = \left({\ds \partial A^{GG}\over\ds \partial m_q}\right)_{\ds m_q=0}
\approx 0.0098\ GeV^{-1}$.

Many - \pl\ \ia s have in principle a small - \dc\
character and may be approximated by the \ef ive t'Hooft \ia\ \ct{tH}
induced by the high - \fq\ part of gluon field \vm\ \ft s, small-size \s s.
In the \s\ \vm\ model \ct{SVZi,Shil} it is expressed by \ct{DKbm}:
\be
\Delta{\cal L}^{(2)}_{inst}= \sum_{i>j}^{i=u,d,s}n_c(k'_ik'_j)
\{\bar q_{iR}q_{iL}\bar q_{jR}q_{jL}
[1+{3\over32}\lambda^{a}_{\ds i}\cdot\lambda^{a}_{j}
(1+3\vec\sigma \cdot \vec\sigma)]+
(R\leftrightarrow L)\}
\la{12.7}\ee
where the coupling
\be
k'_i=\frac{4\pi\rho_C^3}{3}\frac{\pi}{(m^*_i\rho_C)}
\la{InCoupl}\ee
characterizes the \ia\ strength of a \q\  of flavor $i$
with an \s\ and is \pto\ the \s\ volume,
$n_c$ is the \s\ density in the QCD \vm\ related to the \vm\ energy density,
$\varepsilon_{QCD}$, by
$\varepsilon_{QCD}=2n_c$, $n_c={\ds 1\over\ds 16}
<0|{\ds \alpha\over\ds \pi}G^a_{\mu\nu}G^{a\mu\nu}|0>$,
$ \rho_c$ is an \ef ive size of an \s\ in the QCD \vm ,
$q_{R,L}=\frac{1}{2}(1\pm\gamma_{\ds5})q$,
$m^*_i=m_i-{\ds2\over\ds3}\pi^2\rho^2_c<0\mid\bar Q_iQ_i\mid 0>$
is the \ef ive mass
of the \q\ with current mass $m_i$ in physical \vm\ \ct{SVZi}.
An \ef ive mass \tsi\ long-range field correlations in the \s\ \vm .
The term $(R\leftrightarrow L)$ in \re{12.7} corresponds to the \ia\ through
an anti-\s .
The \La\ \re{12.7} is written for $qq$-\ia\ in the $SU_f(2)$ flavor sectors
of the complete $SU_f(3)$ theory.
Selection of $SU_f(2)$ corresponds to the case when two of \q s are exchanged
by a hard \s\ \ft\ and a \q\ of the third \fl\ interacts
with soft \vm\ \cd .
For the $q\bar q$ - system one should change in \re{12.7} \op s of one of \q s
\[
\lambda_{\ds q}^a\rightarrow-\lambda_{\ds\bar q}^{aT},\ \ \ \ \
\vec\sigma_{\ds q}\rightarrow - \vec\sigma_{\ds \bar q}^T.\]

In the recent paper \ct{Klab} D. Klabucar analyzes the octet baryon mass
spectrum in the framework of the MIT \bm\ with \s\ induced \ia . He finds
that the \s\ \con s to \hn\ masses are less than $5\ MeV$ and, therefore,
completely negligible. This conclusion is in strong contradiction with
the results of works \ct{DKbm,ShurRos} and can be traced
to illegal inclusion  of the one - \pl\ part,
$\Delta L^{(1)}_{inst}\propto \bar q_iq_i$ , of the \s\ \ia\
into the calculation of \hn\ masses. This
potentially large \con\ is suppressed then by choosing a very small
value of the package factor $f$, $f=\pi^2n_c\rho^4$
(one thirtieth of a standard value \ct{Shil}).
Then the \con\ of $\Delta L^{(1)}_{inst}$ is at the level of several $MeV$
and that of $\Delta L^{(2)}_{inst}$ is even much less. \Hw , the inclusion
of $\Delta L^{(1)}_{inst}$ being correct in the case of a \q\ in
the background \vm\ field \ct{SVZi} is to be double counting procedure within
the \bm . As it is truly noted in \ct{Klab} within the \nr\ \qm\ \ct{ShurRos}
the one - \pl\ term $\Delta L^{(1)}_{inst}$ is \ef ively \ten\ as a part
of a \cst\ \q\ mass and thus it does not appears explicitly. But the same
occurs within the \rl\ \bm\ \w\ the \cst\ mass results from the bag \bc s:
$m_q\to m_q^{const}=\sqrt{m_q^2+\xi^2R^2}$. The bag \bc s take already
into account \q\ dressing by the \vm\ "medium". That is why in \ct{Klab}
there is no more place for \s s and that is why we do not include the
one - \pl\ term into our considerations.

In the first order in small $u,\ d$ \q\ masses  and \cd\ \di\ we obtain
>from \re{12.7} the \hn\ energy increase resulting from \s s
\be
\Delta E_{\ds inst}=-\frac{E_{\ds inst}^{(os)}(h)}{m^*_0}
\{\sum_{\ds i=u,d}N_{is}m_{\ds i}
-\frac{2\pi^2}{3}\gamma N_{ds}<0| \bar UU |0>\rho_c^2\}
\la{dEinst}\ee
where $N_{is}$ is the number of light - strange scalar di\q s in a \hn , $h$,
$m^*_0$ is an \ef ive mass of a \q\ with a zero \cu\ mass and
$E^{os}_{inst}(h)$ is the \s\ correction for these di\q s
calculated with the static spherical cavity \wf s
\be
E^{os}_{inst}(h)=-<h| \Delta L^{os}_{inst} |h>.
\la{Einst}\ee

The values of $E^{os}_{inst}(h)$ for members of \hn\ \mlp s are the
following:
\ba
&&E^{os}_{inst}(\pi)=0, \ \ \ \ \
E^{os}_{inst}(K)=-{\ds \lambda^0_{0s}\over\ds R^3},    \nn\\
&&E^{os}_{inst}(N)=0, \ \ \ \ \
E^{os}_{inst}(\Sigma)=E^{os}_{inst}(\Xi)=
-{\ds 1 \over\ds 2R^3}\left({\ds 3\over\ds 2}\lambda^0_{0s}+
{\ds 1\over\ds 2}\lambda^1_{0s}\right),
\la{InstInt}\ea
\w\ $\lambda^l_{i,j}=n_ck'_ik'_jI^l_{i,j}$ with $l$ for spin of a di\q\
and \il s are
\ba
I^0_{i,j}&=&{\ds 3\over\ds 4\pi}N^2_{\ds i}N^2_{\ds j}\int^1_0dx\ x^2
\left[\sqrt{1+{m_{\ds i}R\over\omega_i}}\sqrt{1+{m_{\ds j}R\over\omega_j}}
j_0(\xi_{\ds i}x)j_0(\xi_{\ds j}x)+ \right. \nn\\
&+&\left.
\sqrt{1-{m_{\ds i}R\over\omega_i}}\sqrt{1-{m_{\ds j}R\over\omega_j}}
j_1(\xi_{\ds i}x)j_1(\xi_{\ds j}x)\right]^2,
\nn\ea
\ba
I^1_{i,j}&=&-{\ds 1\over\ds 4\pi}N^2_{\ds i}N^2_{\ds j}\int^1_0dx\ x^2
\left[\sqrt{1+{m_{\ds i}R\over\omega_i}}\sqrt{1-{m_{\ds j}R\over\omega_j}}
j_0(\xi_{\ds i}x)j_1(\xi_{\ds j}x)- \right. \nn\\
&-&\left.
\sqrt{1-{m_{\ds i}R\over\omega_i}}\sqrt{1+{m_{\ds j}R\over\omega_j}}
j_1(\xi_{\ds i}x)j_0(\xi_{\ds j}x)\right]^2,
\nn\ea
$N_{\ds i}$ is the normalization of the
\wf\ of a \q\ with mass $m_i$:
\ba
N_{i}^{-2}=
j^2_0(\xi_i)[2\omega_i(\omega_i-1)+m_iR]/\omega_i(\omega_i-m_iR).
\nn\ea
$\xi_i$ is a root of the equation \re{egval}. We should note that the \con\
to a vector di\q\ results from the inequality of \q\ masses and is very small
as compared to the scalar di\q\ \il\ even on the scale of a \sg\ \q\ mass.

Due to the determinant character of \s\ induced \q\ \ia\ \re{12.7},
for members of the meson vector nonet and baryon decouplet the corrections
\re{Einst} are equal to zero. This selection rule comes from
the fact that the \s\ mechanism of \ia\ within a \hn\ takes place only
if two \q s are in the state with zero total spin (plus color spin),
the scalar di\q .

The results of calculations are presented in Table I.
The numbers in Table I correspond to the \pr s of the QCD \vm :
\ba
&&<0| \bar UU |0> = -(221\ MeV)^3,                             \nn\\
&&<0| {\ds \alpha_s\over\ds \pi}G^2 |0>=0.031\ GeV^4,
\ \ \ \ \rho_c^2=1\ GeV^{-2},
\la{par}\ea
the \di s        of  $u$ and $d$ \q\ masses ($m_u = 5\ MeV$)
\be
m_d-m_u= 3.5\ \mbox{MeV},
\la{Mdu}\ee
and their \cd s
\be
\gamma=\frac{<\bar DD> - <\bar UU>}{<\bar UU>}=-2\cdot10^{-3}.
\la{CondDif}\ee
In our calculations we take the values $m_{\ds s}=200\ MeV$,
$\alpha_{\ds s}=0.4$ and
$\delta=\frac{\ds <\bar SS> - <\bar UU>}{\ds <\bar UU>}=-0.1$.
We should note that the reduction of $\Delta m_q$, as compared with \ct{BTh},
is due to the center - of - mass motion \ef s \re{DMtot}.

We use the value of the \q\ \cd\ which agrees with the standard one \ct{RRYa}:
$<0| \bar UU |0> = -(225 \pm 25\ MeV)^3$. The value of the gluon \cd\
is close to a recent estimation extracted from the two loop fit of charmonium
data \ct{Broad}:
$<0| {\ds \alpha_s\over\ds \pi}G^2 |0>=0.021\ GeV^4$
which is almost twice as the standard one \ct{RRYa}. The \ia\ with the
\cd s resemble the one - \pl\ \con\ to a \q\ mass due to long-range \ft s
of \vm\ medium and for the \p\ state this increase is approximately equal to
$\Delta m^{vac}_q \approx 270\  MeV$.
On the other hand, the value of the \s\ \q -\q\ \ia\ strength is sensitive
to the ratio of \q\ and gluon \cd s, \re{12.7}, \re{InCoupl}, and provides a
large negative \con\ to the \p\ energy, $\Delta E_{inst}^{00}\approx -210\
MeV$.

The value of $\rho_c^2$ that we use in this paper is slightly less than
obtained in the \s\ liquid model \ct{Shil,DPil}. It leads to a lower value of
an \ef ive (chiral) mass \pr\
$m^*_0\approx 70\ MeV$. This causes two \ef s: large \s\ \con s to the
EMD of baryon the octet due to the
$1/m_0^*$ dependence of these \di s and a more stronger suppression of
\s\ \ia\ in the light - strange di\q : $m_0^*/m^*_s\approx 0.3$
as compared with $m_0^*/m^*_s\approx 0.6$ in \ct{Shil}.
Small values of $\rho_c^2$ and $m_0^*$ are characteristic of the
chiral phase in the framework of the confining QCD \vm\ model developed
in \ct{Sim}.

Another important \vm\ \pr\ is a packing fraction characterizing the
diluteness of \s\ \vm . With our choice of \pr s it is quite small:
\be
f=(2n_c){\ds \pi^2\rho^4_c\over\ds 2}\approx 1/50
\la{f}\ee
and justifies the one - \s\ approximation used. The value of $f$ that we use
corresponds exactly to the one obtained in the Monte Carlo lattice
calculations \ct{DiGif}.

We now turn to the discussion of EMD.
There are two exceptional EMD combinations which depend only on the \elm\
term $\Delta E_{em}$ thus being sensitive only to the bag radius. They are
the $I=2$ part of the $\Sigma -$ and the $\pi$ mass differences:
$\Sigma^+ + \Sigma^- - 2\Sigma^0 = 1.71\pm0.15\ MeV$,
$\pi^\pm - \pi^0=4.59\pm0.05\ MeV$ (this is valid for
$\rho^\pm - \rho^0$, too, but its \ex al value is not well defined). The
first one is satisfied in the range of bag radii $R= 5\sim6\ GeV^{-1}$, which
confirms a good and self - consistent description of the mass scales of the
baryon octet and splittings within it. From Table I we
see that the bag stability radius, $R= 5.6\ GeV^{-1}$, belongs to this interval
and $\Sigma^+ + \Sigma^- - 2\Sigma^0 = 1.71\ MeV$ in fine agreement with \ex al
value.
As to the pion, due to large negative \s\ and c.m. energy \con s, it has no
radius of bag stability. This is a signal of the Goldstone nature of the pion
in our model.
As it has previously been pointed out \ct{DKbm},
the \ef s of relativism and multiparticle \str\ of the pion \wf\
are urgently necessary to describe the pion within the \bm .

Given a typical bag radius and due to the absence of \s\ \con , the $p - n$
mass \di\ ($p-n = 1.3\ MeV$) is
mainly defined by the sum of the \elm\ energy term, $\Delta E_{em}$, and
the kinetic energy term, $\Delta E_{kin}$. The latter directly
depends on the $u - d$ \q\ mass \di . We could
obtain the \ex al value precisely by fitting $\Delta m_q$ with an accuracy of
$0.01\ MeV$. \Hw , we think that the calculations with such a high
accuracy can not be made within the \bm\ approach. Within some
uncertainties in the definition of the model \pr s, the value of $\Delta m_q$
is
grouped around $3.5\ MeV$.

Usually, in the bag model there is the problem with the description of the
$I=1$ part of octet splitting. In fact, the bag radius and $\Delta m_q$ fixed
as above, it is impossible to saturate the Coleman - Glashow relation (CG),
$p - n + \Xi^0 - \Xi^- = \Sigma^+ - \Sigma^-$, only with the kinetic energy,
$\Delta E_{kin}$, and \elm\ energy, $\Delta E_{em}$,
\con s.
That is, $CG^{theor}_0=\Delta E_{kin}^{CG}+\Delta E_{em}^{CG}+...\approx
-4.5\ MeV$ while the \ex al value is about $CG^{ex}\approx -8\ MeV$.
It is important to stress that the color - magnetic energy \con\ could not
save the situation with CG  even for the large constant $\alpha_s^{MIT}=2.2$.

As it has been noted in \ct{DKEM1,YWL} the $I=1$ part of the $\Sigma -$ and
$\Xi -$ mass \di s is essentially due to the \s\ \con\ that is \pto\ the number
of \sg\ \q s. It reproduces the term introduced \ph ycally in \ct{Desh}.
It is important that these \spl s are of the same order as the
$u-,\ d-$ \q\ mass \di s. Then from \re{dEinst} one has:
$\Delta E_{inst}^\Sigma={\ds E^{0s}_{inst}\over\ds m^*_0}\Delta m_q,$
and for typical values for
$E_{inst}^\Sigma \approx -70\ MeV$ it follows that the \ef ive \q\ mass
$m^*_0$ is of the
same order as $E_{inst}^\Sigma$. This ratio requires quite a small value for
$m^*_0\sim<\bar QQ>\rho_c^2$ and a large value for a gluon \cd\
$E_{inst}\sim<0| {\ds \alpha_s\over\ds \pi}G^2 |0>/<\bar QQ>^2$, \re{par}.
The Coleman - Glashow relation is satisfied by each \con\
separately because, as noted above, the bag
radii for $N,\ \Sigma,\ \Xi$ are well equal. As to the absolute value of the
left and right sides of this relation, the \s\ \con\ is very important.
 From Table I we find $p - n + \Xi^0 - \Xi^- = -8.30\ MeV$
$(-7.7\pm0.6\ MeV)_{exp}$ and $\Sigma^+ - \Sigma^- = -8.06\ MeV$
$(-8.07\pm0.09\ MeV)$ in excellent agreement with \ex\ values.

Thus, it is shown that the \s\ plays the key role in the saturation of the CG
relation between octet baryon states. In this case, as in the case of the
dynamical
explanation of the Okubo - Zweig - Iizuka rule \ct{GI}, the gluon exchange
contributions are very small and, therefore, from the magnitudes of these
\ef s we can clearly judge on the strength of the \s\ induced \ia .

The \con s related to the \cd\ \di\ are not large, act opposite
to the first terms in \re{dEvac}, \re{dEinst} and are poor fixed from \hn\ mass
\di s. From our analysis we can define only the lower bound of $\gamma$:
$0>\gamma>-0.003$. With a precision of the model and data it is difficult
to expect for more.

We have compared the results using linear and quadratic bag model formulae.
As a rule, the center - of - mass corrections lead to  larger bag radii,
additional corrections to $\Delta E_{kin}$ which partially contribute
to the CG relation. The main \ef\ of the center of mass motion corrections
on \pr s reduce
$\Delta m_q$ approximately by $0.5\ MeV$.

 From Table I we see that the \ia\ induced by \s s gives an essential
\con\ to the isotopic mass \di s of \hn s belonging to a baryon octet and \ps\
mesons. As expected, the \qt\ agreement with \ex al values for the \ps\ octet
is not entirely satisfactory. As noted above, in the problem of the masses and
their splittings of the \ps\ octet it is necessary to \tia\ the higher orders
in the \s\ \ia .
Another problem is the \di\ of vector strange mesons $K^{*+}-K^{*0}$.
This discrepancy between the \th\ prediction and the values from Particle Data
looks strange because we do not see any large \con\ to this \di . We
hope that more precise \ex s on determining the \elm\ \di s of vector
resonances will clear up the situation.

At last, we would like to say a few words about other approaches.
In \ct{BrownEMD,IoffeEMD,Kissl}, the problem concerned
has been discussed within the QCD sum rule method.  There,
the important \con\ of
the interaction with small-size instantons that dominates in the
short wavelength region of vacuum fluctuations has been missed.
\Hw , in our opinion, this interaction is of principal importance. It violates
the \q\ additivity, specifies spin-spin splitting in the hadron mass spectrum
and
determines the mixing angles in the hadron $SU(3)_f$ multiplets.
In \ct{Shurpi,NucSR}, it has been shown that the consideration of the QCD  sum
rules
for the pion and proton confirms the fundamental role of instanton interaction
on
which the model is based. This conclusion is also proved
in Lattice QCD calculations \ct{shurak}. Another problem of the QCD sum rules
method is to \tia\ the $\Sigma^0 - \Lambda$ mixing \ct{IoffeEMD}, the \ef\
of which is negligible within the \qm\ \ct{Frank}.

In summary, we conclude from our results for isospin mass \hn\ differences
that $m_d-m_u = 3.5 MeV$ and
$<\bar DD> - <\bar UU> = -( 0\sim 3)\cdot10^{-3}<\bar UU> $.
It would be interesting to consider the $D$ and $D^*$ isospin mass \di s
in the framework of the \q\ model with QCD \vm\ induced \ia .

The author is very thankful to Professor A.W. Thomas and members of the
theoretical
seminar of Adelaide University and N.N. Achasov, S.B. Gerasimov, N.I. Kochelev
for stimulating discussions. He is also thankful to Dr. E. Rodionov for
collaboration at the beginning stage of this work.
\vskip 7mm

\bb{99}
\bi{HatsudaEMD} T. Hatsuda, H. Hogaasen, M. Prakash,
{\it Phys. Rev.} {\bf C42} (1990) 2212;
{\it Phys. Rev. Lett.} {\bf 66} (1991) 2851;
\bi{BrownEMD} C. Adami, G.E. Brown {\it Z. Phys.}{\bf A340} (1991) 93.
\bi{IoffeEMD} G. Adami, E. Drukarev, B.L. Ioffe
{\it Phys. Rev.} {\bf D48} (1993) 2304.
\bi{EletIof} V.L. Eletsky, B.L. Ioffe {\it Phys. Rev.} {\bf D48}
(1993) 1587.
\bi{NasonEMD} E. Gabrielli, P. Nason {\it CERN preprint}
CERN-TH.6857/93 (1993).
\bi{Kissl} K.-C. Yang, W.-Y.P. Hwang, E.M. Henley, and L.S. Kisslinger
{\it Phys. Rev.} {\bf D48} 3001 (1993).
\bi{JinNiel} X. Jin, M. Nielsen, J. Pasupathy {\it preprint}
DOE/ER/40762-027; UMPP 94-082 (1994).
\bi{DPG} Review of Particle Properties {\it Phys. Rev. }{\bf D45} (1992) 1.
\bi{CSBrev} G.A. Miller, B.M.K. Nefkens, I. Slaus
{\it Phys. Rep. }{\bf 194} (1990) 1.
\bi{Bo} P. N. Bogoliubov {\it Ann.\ Inst.\ H.\ Poincare} {\bf 8} (1967) 163.
\bi{mit2} A. Chodos, R. L. Jaffe, K. Johnson, and C. B. Thorn
{\it Phys.\ Rev.\ }{\bf D10} (1974) 2599.
\bi{Viollier} G.V. Schreiber, R.D. Viollier
{\it Ann. of Phys.} {\bf 215} 277 (1992) and references therein.
\bi{Desh} N.G.Deshpande et al {\it Phys.\ Rev.\ }{\bf D15} (1977) 1885;
L.P.Singh {\it Phys.\ Rev.\ }{\bf D22} (1980) 2224.
\bi{BTh}  R.P. Bickershtaff, A.W. Thomas {\it Phys.\ Rev.\ }{\bf D25} (1982)
1869.
\bi{Leutw} See for a recent review, H. Leutwyller {\it Bern University
preprint} {\bf BUTP-94/8} (1994).
\bi{DKEM1} A. E. Dorokhov, N. I. Kochelev {\it Z. Phys.}{\bf C37} (1988) 377.
\bi{YWL} G. Yang, L. Wang, H.C. Liu {\it Z. Phys.}{\bf C26} (1984) 77.
\bi{DKbm} A. E. Dorokhov, N. I. Kochelev
{\it Sov. J. Nucl. Phys.} {\bf 52} (1990) 135 (214);\\
In Proc. Int. Conf. {\it Quarks-86, Tbilisi}, p. 392, Moscow, 1986;\\
A. E. Dorokhov, N. I. Kochelev, Yu. A. Zubov {\it Sov. J. Part. Nucl. Phys.}
{\bf 23} (1992) 522 (1192).
\bi{Koch} N. I. Kochelev {\it Sov. J. Nucl. Phys.}{\bf 41} (1985) 291.
\bi{ShurRos} E. V. Shuryak, J. Rosner {\it Phys. Lett.} {\bf 218B} (1989) 72.
\bi{Hanss} T. H. Hansson {\it Nucl. Phys.} {\bf B249} (1986) 742.
\bi{SVZ} M.A.Shifman, A.I.Vainstein, V.I.Zakharov {\it Nucl.\ Phys.\ }
{\bf B147} (1979) 385.
\bi{RRYa} For a review, see L.J. Reinders, H. Rubinustein, S. Yazaki
{\it Phys.\ Rep.\ }{\bf 127} (1985) 1.
\bi{DJ} J. Donoghue, K. Johnson, {\it Phys.\ Rev.\ }{\bf D21} (1980) 1975.
\bi{tH} 't Hooft {\it Phys. Rev.} {\bf D14} (1976) 3432;
{\it Phys. Reports} {\bf 142} (1986) 357.
\bi{SVZi} M.A.Shifman, A.I.Vainstein, V.I.Zakharov {\it Nucl.\ Phys.\ }
{\bf B163} (1980) 45.
\bi{Shil} E. V. Shuryak, {\it Phys.\ Reports} {\bf115} (1984) 151.
\bi{Klab} D. Klabucar {\it Phys. Rev.} {\bf D49} (1994) 1506.
\bi{DPil} D.I. Dyakonov, V.Yu. Petrov {\it Nucl.\ Phys.\ }
{\bf B272} (1986) 457.
\bi{Shurpi} E. V. Shuryak, {\it Nucl. Phys.} {\bf B214} (1983) 237.
\bi{Broad} D.J. Broadhurst et. al. {Preprint} OUT-4102-49, March, 1994.
\bi{Sim} Yu.A. Simonov {ITEP preprint} {\bf 83-93} Moscow, 1993.
\bi{DiGif} M. Campostrini, A. Di Giacomo, G. Mussardo {\it Z.Phys.} {\bf C25}
(1984) 173.
\bi{GI} B.V. Geshkenbein, B.L. Ioffe {\it Nucl.\ Phys.\ } {\bf B166} (1980)
340.
\bi{NucSR} A.E.Dorokhov, N.I.Kochelev {\it Z.Phys.} {\bf C46} (1990) 281.
\bi{shurak} E. V. Shuryak, {\it Rev.\ Mod. Phys.} {\bf65} (1993) 1.
\bi{Frank} J. Franklin et. al. {\it Phys. Rev.} {\bf D24} (1981) 2910.
\eb
\newpage
\begin{table}[h]
Table I. The electromagnetic mass differences of Hadrons (MeV), \re{DMtot},
($\Delta M_{exp}$ from \ct{DPG}). The \pr s used are:
\ba
&&<0| \bar UU |0> = -(221\ MeV)^3,
<0| {\ds \alpha_s\over\ds \pi}G^2 |0>=0.031\ GeV^4,
\rho_c^2=1\ GeV^{-2},
\nn\\
&&m_d-m_u= 2.5\ \mbox{MeV}, m_{\ds s}=200\ MeV, \alpha_{\ds s}=0.4
\nn\ea
\begin{center}
\small
\vskip 2mm
\hskip -0.cm
      \begin{tabular}     {|c| c|  c|   c|   c|    c|    c|    c|     c|
c|     l|} 	\hline \hline
 Particles & R & M &$\Delta M_{kin}$&$\Delta M_{EM}$&$\Delta M_{gl}$&
                                                $\Delta M_{vac}$&$\Delta
M_{inst}$&

$\Delta M_{tot}$&$\Delta M_{exp}$  \\ \hline  \hline
        \hline
 P-n                         &5.60& 940  &-1.06& 0.55& 0.04&-0.15&  0
&-1.23&-1.2933$\pm$ 0  002                   \\
$\Sigma ^+ - \Sigma ^0$      &5.40&1230  &-1.19& 0.55&-0.01&-0.13&-1.49
&-3.18&-3.18  $\pm$0.17                      \\
$\Sigma ^0 - \Sigma ^-$      &    &      &-1.19&-1.16&-0.01&-0.13&-1.49
&-4.89&-4.89  $\pm$0.08                      \\
$\Xi ^{0}-\Xi ^{-}$          &5.30&1330  &-1.22&-1.21&-0.07&-0.13&-3.16
&-7.07&-6.4   $\pm$0.6                       \\
$\Delta ^{++}-\Delta ^{0}$   &6.40&1240  &-2.56& 2.51& 0.08&-0.48&  0
&-1.75&-2.70  $\pm$0.30                      \\
$\Delta^{+}-\Delta ^{0}$     &    &      &-1.28& 0.48& 0.04&-0.24&  0   &-1.61&
                                      \\
$\Delta ^0-\Delta ^{-}$      &    &      &-1.28&-0.97& 0.04&-0.24&  0   &-2.90&
                                      \\
$\Sigma ^{*+} - \Sigma ^{*-}$&6.65&1380  &-2.66&-0.50& 0.08&-0.56&  0
&-4.89&-4.4   $\pm$0.5                       \\
$\Sigma ^{*0} - \Sigma ^{*-}$&    &      &-1.33&-0.94& 0.04&-0.28&  0
&-3.03&-3.5   $\pm$1.2                       \\
$\Xi ^{*0}-\Xi ^{*-}$        &6.70&1510  &-1.37&-0.96& 0.03&-0.29&  0
&-3.18&-3.2   $\pm$0.6                       \\
$K^+ -K^0$                   &5.70&700   &-0.87& 0.56&-0.11&-0.16&-3.34
&-5.40&-4.024 $\pm$0.032                      \\
$\rho ^{\pm}-\rho^0$         &6.00&780   & 0.00& 0.77&  0  &  0  &  0   &
0.75&-0.3   $\pm$2.2                       \\
$K^{*+} -K^{*0}$             &6.00&890   &-1.07& 0.53& 0.03&-0.20&  0
&-1.19&-6.7   $\pm$1.2                        \\
     \hline  \hline
	\end{tabular}
\end{center}
\la{tablem}
\end{table}
\ed